\documentclass[letter,12pt]{article}
\usepackage{graphicx}
\begin{document}

\begin{flushleft}
{\sf {\Large Long-Distance Distribution of Time-Bin Entanglement Generated in a Cooled Fiber}} 
\end{flushleft}

\begin{flushleft}
Hiroki Takesue$^{1,2}$\\
$^1$NTT Basic Research Laboratories, NTT Corporation\\
3-1 Morinosato Wakamiya, Atsugi, Kanagawa, 243-0198, Japan\\
$^2$CREST, Japan Science and Technology Agency\\
4-1-8 Honcho, Kawaguchi, Saitama, 332-0012, Japan\\
\today
\end{flushleft}

\begin{flushleft}
Abstract
\end{flushleft}
This paper reports the first demonstration of the generation and distribution of entangled photon pairs in the 1.5-$\mu$m band using spontaneous four-wave mixing in a cooled fiber. 
Noise photons induced by spontaneous Raman scattering were suppressed by cooling a dispersion shifted fiber with liquid nitrogen, which resulted in a significant improvement in the visibility of two-photon interference. By using this scheme, time-bin entangled qubits were successfully distributed over 60 km of optical fiber with a visibility of 76\%, which was obtained without removing accidental coincidences. 

\newpage
The entangled states of quantum particles constitute the quintessential feature of quantum mechanics because they highlight its non-locality most vividly \cite{epr}. Moreover, entanglements form the basis of quantum information, and facilitate such applications as quantum key distribution (QKD) \cite{ekert,bbm92}, quantum teleportation \cite{bennett3}, and quantum repeaters \cite{briegel}. Of the many forms of entanglement, entangled photons are important because they are suitable for distributing quantum information over long distances. 
Although several good sources are available in the short wavelength band \cite{kwiat1,kwiat2}, an entangled photon-pair source in the 1.5-$\mu$m band is needed if we are to realize quantum information systems over optical fiber networks. 
Recently, spontaneous four-wave mixing (SFWM) in a dispersion shifted fiber (DSF) has been drawing attention as a promising method for generating entanglement in the 1.5-$\mu$m band \cite{li,takesue,takesue2}. 
Polarization \cite{li,takesue} and time-bin entangled photons \cite{takesue2} have already been generated and successfully distributed over optical fibers \cite{takesue,takesue2,li2}. The merit of this scheme is the good coupling efficiency it provides between nonlinear medium (i.e. a DSF) and transmission fibers. 

Despite the series of successful experiments described above, a serious problem has been reported regarding the fiber-based photon-pair source, namely the existence of noise photons generated by spontaneous Raman scattering \cite{li,takesue,takesue2,li3,kyo}. The pump light used for the SFWM process also works as the pump for the spontaneous Raman scattering process, by which Stokes and anti-Stokes photons are generated in wavelength bands longer and shorter than the pump wavelength, respectively. 
Consequently, accidental coincidences are caused by the Stokes and anti-Stokes photons whose wavelengths coincide with those of the idler and signal channels of photon pairs generated by SFWM. In the two-photon interference measurement, such accidental coincidences seriously degrade the visibility. As a result, the visibilities reported in the previous experiments were obtained after subtracting accidental coincidences \cite{li,takesue,takesue2,li2}. This large number of noise photons prevented the fiber-based photon-pair source from being applied to quantum information experiments such as QKD, because accidental coincidences result in a large bit error rate. Thus, the suppression of accidental coincidences caused by spontaneous Raman scattering is a very important issue as regards making the fiber-based photon-pair source useful in real quantum information systems.  

To solve this problem, Inoue and I recently demonstrated that noise photons caused by spontaneous Raman scattering were suppressed by cooling a DSF with liquid nitrogen \cite{takesue3}. We observed a significant enhancement of the quantum correlation characteristics in a time-correlation measurement. However, entangled states have yet to be generated using a cooled fiber, and so the improvement in the visibility has not been confirmed directly in a degree-of-entanglement measurement. 

In this paper, I report the first experiment on time-bin entanglement generation in a DSF cooled by liquid nitrogen. With an average photon number per pulse of $\sim$0.06, a significant improvement in the visibility from 65\% to 80\% was achieved by cooling the DSF, without subtracting accidental coincidences. In addition, time-bin entangled qubits were successfully distributed over 60 km (30 km x 2) of optical fiber with a visibility of 76\%. This distance exceeds the previous record for the long-distance distribution of entanglement set by a group from Geneva University (25 km x 2), in which they used a lithium triborate crystal as a nonlinear medium \cite{marc}.

Spontaneous Raman scattering is a process in which a spontaneous photon is generated by a nonlinear interaction between a pump photon and a phonon. The numbers of Stokes, $n_s$, and anti-Stokes photons, $n_{as}$ as a function of temperature $T$ are expressed as \cite{takesue3}
\begin{eqnarray}
n_s (T) &=& \frac{g L e^{-\alpha L}}{1-\exp\left(-\frac{h \nu}{k_B T}\right)}, 
\label{sss}\\
n_{as} (T) &=& \frac{g L e^{-\alpha L}}{\exp\left(\frac{h \nu}{k_B T}\right)-1}, 
\label{aaa}
\end{eqnarray}
where $g$, $\alpha$, $L$, $h$, $\nu$ and $k_B$ are a gain coefficient proportional to the pump power, a fiber loss coefficient, the fiber length, Planck constant, photon energy, and Boltzman constant, respectively. It is apparent that we can reduce the number of spontaneous Raman photons by lowering the fiber temperature. 
If the fiber loss coefficient is independent of temperature, the ratio of the Stokes photon number at liquid nitrogen (77 K) and room temperature (293 K) with the same pump power is calculated to be $n_s (77)/n_s(293)=0.29$. For anti-Stokes photon numbers it is $n_{as}(77)/n_{as}(293)=0.24$. Here, I assumed that $\nu=400$ GHz, which corresponds to the frequency difference between the pump and signal/idler channels in the experiment. Thus, the number of noise photons is expected to be reduced by cooling the fiber with liquid nitrogen.

\begin{figure}[thb]

\centerline{\includegraphics[width=\linewidth]{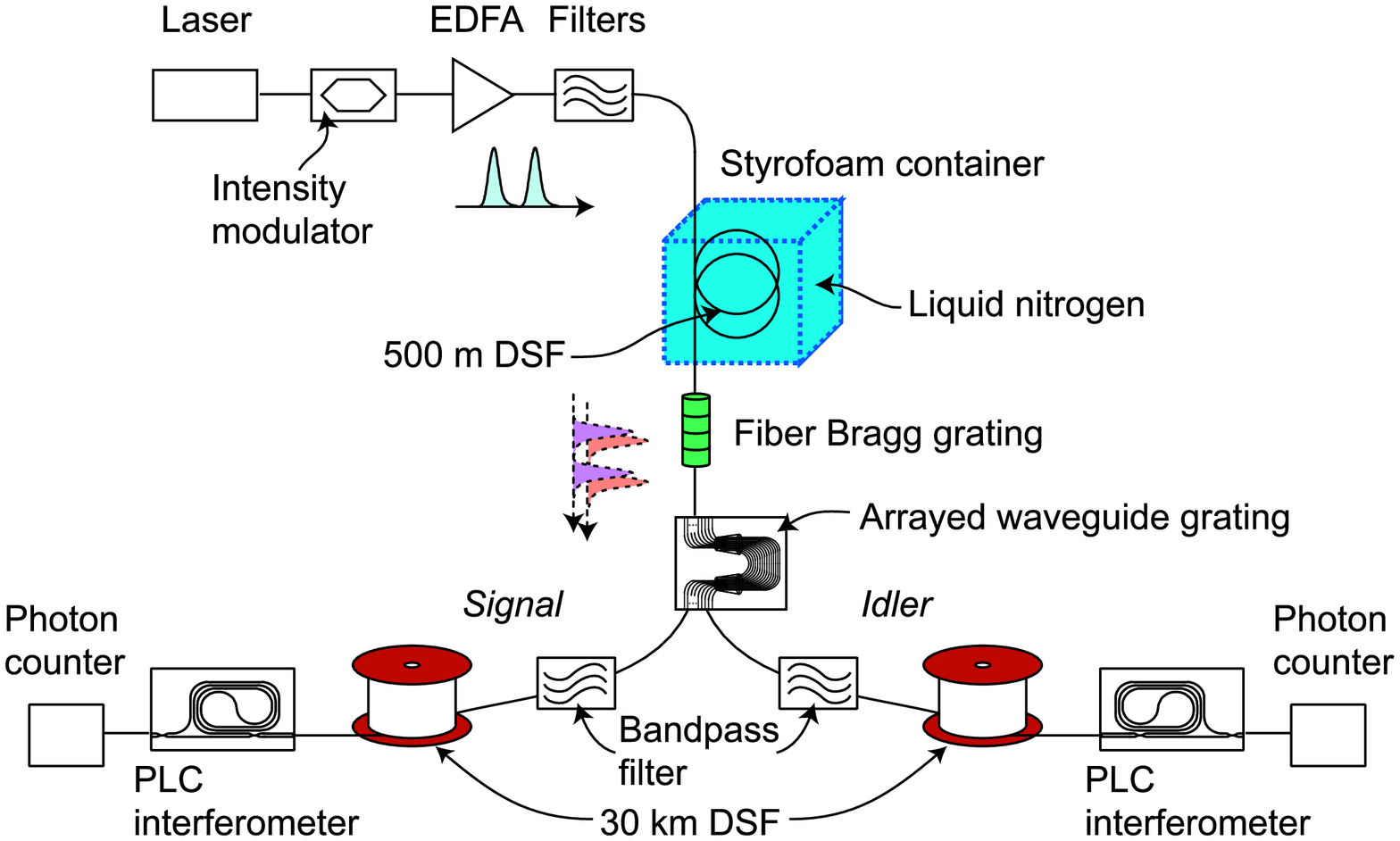}}

\caption{Experimental setup. }
\label{es}

\end{figure}

Figure \ref{es} shows the experimental setup. 
A continuous lightwave with a wavelength of 1551.1 nm from an external-cavity diode laser is modulated into double pulses with a LiNbO$_3$ intensity modulator. The pulse width and interval are 100 ps and 1 ns, respectively. The coherence time of the continuous lightwave is $\sim$10 $\mu$s. The double pulse is amplified by an erbium-doped fiber amplifier (EDFA), and launched into a 500-m DSF after passing through optical filters to eliminate amplified spontaneous emission noise from the EDFA. The DSF is formed into a loose coil about $\sim$30 cm in diameter without a bobbin to reduce the bending stress, and then placed in a Styrofoam container filled with liquid nitrogen. The zero-dispersion wavelength of the DSF was 1551.1 nm. 
In the DSF, the double pulses work as a pump and generate time-correlated photon pairs through SFWM. The pump, signal and idler frequencies, $f_p$, $f_s$ and $f_i$, respectively, have the following energy conservation relationship: $2 f_p = f_s + f_i$. 
As a result, the following time-bin entangled state is obtained at the output of the DSF \cite{brendel}.
\begin{equation}
|\Psi\rangle = \frac{1}{\sqrt{2}} \left(|1\rangle_s |1\rangle_i + e^{i\phi} |2\rangle_s |2\rangle_i \right)
\end{equation}
Here, $|k\rangle_x$ represents a state in which there is a photon in a time slot $k$ in a mode $x$, signal ($s$) or idler ($i$). $\phi$ is a relative phase term that is equal to $2 \phi_p$, where $\phi_p$ is the phase difference between two pump pulses, and is stably fixed because of the long coherent time of the laser output. 
The output light from the DSF is input into a fiber Bragg grating (FBG) to suppress pump photons, and launched into an arrayed waveguide grating (AWG) to separate the signal and idler channels. AWG output ports with peak frequencies of $+400$ and $-400$ GHz from the pump photon frequency are used for the signal and idler, respectively. The 3-dB bandwidths of the signal and idler channels are both 25 GHz ($\simeq$ 0.2 nm). Then the signal and idler photons are launched into optical bandpass filters to further suppress the pump photons.

The photons output from each bandpass filter are transmitted over a 30-km DSF with a loss of 0.2 dB/km and then input into a 1-bit delayed Mach-Zehnder interferometer fabricated using planar lightwave circuit (PLC) technology \cite{honjo,takesue2}. 
A state $|k\rangle_x$ is converted as follows by the interferometer.
\begin{equation}
|k\rangle_x \to \frac{1}{2} \left(|k,a\rangle_x - |k,b\rangle_x + e^{i \theta_x} |k+1,a\rangle_x + e^{i \theta_x}|k+1,b \rangle_x \right) \label{conv}
\end{equation}
Here, $a$ and $b$ denote two output ports of the interferometer. $\theta_x$ is the phase difference between the two paths of the interferometer for channel $x$, and can be tuned by changing the temperature. 
As a result, the time-bin entangled state $|\Psi\rangle$ is converted to 
\begin{eqnarray}
|\Psi \rangle &\to& \frac{1}{4 \sqrt{2}} \left\{|1,a\rangle_s |1, a\rangle_i - |1,a\rangle_s |1,b\rangle_i \right. \nonumber \\
& & - |1,b\rangle_s |1, a\rangle_i + |1,b\rangle_s |1,b\rangle_i \nonumber \\
& & + (e^{i\theta}+e^{i\phi})|2,a\rangle_s |2, a\rangle_i + (e^{i\theta}-e^{i\phi})|2,a\rangle_s |2,b\rangle_i\nonumber \\
& & +(e^{i\theta}-e^{i\phi})|2,b\rangle_s |2, a\rangle_i + (e^{i\theta}+e^{i\phi})|2,b\rangle_s |2,b\rangle_i \nonumber \\
& & e^{i (\phi+\theta)} |3,a\rangle_s |3, a\rangle_i +e^{i (\phi+\theta)}|3,a\rangle_s |3,b\rangle_i \nonumber \\
& & + e^{i (\phi+\theta)}|3,b\rangle_s |3,a\rangle_i + e^{i (\phi+\theta)}|3,b\rangle_s |3,b\rangle_i \nonumber \\
& & \left. +\cdots
\right\}, \label{after}
\end{eqnarray}
where $\theta=\theta_s +\theta_i$ and 16 non-coincident terms in the parentheses are not shown because they are not observed in a coincidence measurement. We can observe a two-photon interference fringe by changing $\theta$ and measuring the coincidence counts in the second time slot. 

Port $a$ of each interferometer is connected to a photon counter based on an InGaAs avalanche photodiode operated in a gated mode with a 4-MHz gate frequency. The electric signals from the photon counter for the signal and the idler are input into a time interval analyzer as a start and stop pulse, respectively. The losses of the signal and idler channels including the excess losses of the interferometers are both approximately 8 dB. The quantum efficiencies and dark count rate per gate are 8\% and $4 \times 10^{-5}$ for the signal, and 7\% and $5 \times 10^{-5}$ for the idler, respectively. 

This experiment uses two detectors connected to port $a$ of the interferometers, so only the fifth term in parentheses in Eq. (\ref{after}) is observed in a two-photon interference measurement. This means that a time-bin entangled photon pair is detected with a probability of 1/8 when a constructive interference occurs (i.e. $\theta=\phi$). When the average number of correlated photon pairs per pulse is $\mu_c$ (which means that the average number per time-bin qubit is $2 \mu_c$), the count rate of correlated events in a constructive interference, $R_c$, is expressed as
\begin{equation}
R_c = \frac{\mu_c}{4} \alpha_s \alpha_i, \label{cor}
\end{equation}
where $\alpha_s$ and $\alpha_i$ denote transmittances for the signal and idler channels including the quantum efficiency of the photon counters, respectively. 
On the other hand, the accidental coincidence rate $R_{acc}$ is given by
\begin{equation}
R_{acc} = \left(\frac{\mu_s \alpha_s}{2}+d_s\right)\cdot\left(\frac{\mu_i \alpha_i}{2} + d_i\right)\simeq \frac{\mu_s \mu_i}{4} \alpha_s \alpha_i, \label{acc}
\end{equation}
where $\mu_x$ and $d_x$ denote the average number of photons per pulse and the dark count rate of the detector for channel $x$ with $x=s,i$, respectively. 
If the average number of noise photons per pulse for channel $x$ is given by $\mu_{nx}$, $\mu_x$ is expressed as
\begin{equation}
\mu_x = \mu_c + \mu_{nx}.
\end{equation}
A count rate of $R_c+R_{acc}$ and $R_{acc}$ is observed at the maximum and minimum points of a two-photon interference fringe.  
Therefore, the visibility $V$ is expressed as
\begin{equation}
V=\frac{R_c}{R_c+2R_{acc}}\simeq \frac{\mu_c}{\mu_c + 2\mu_s \mu_n}. \label{v}
\end{equation}

\begin{figure}[h]

\centerline{\includegraphics[width=.75\linewidth]{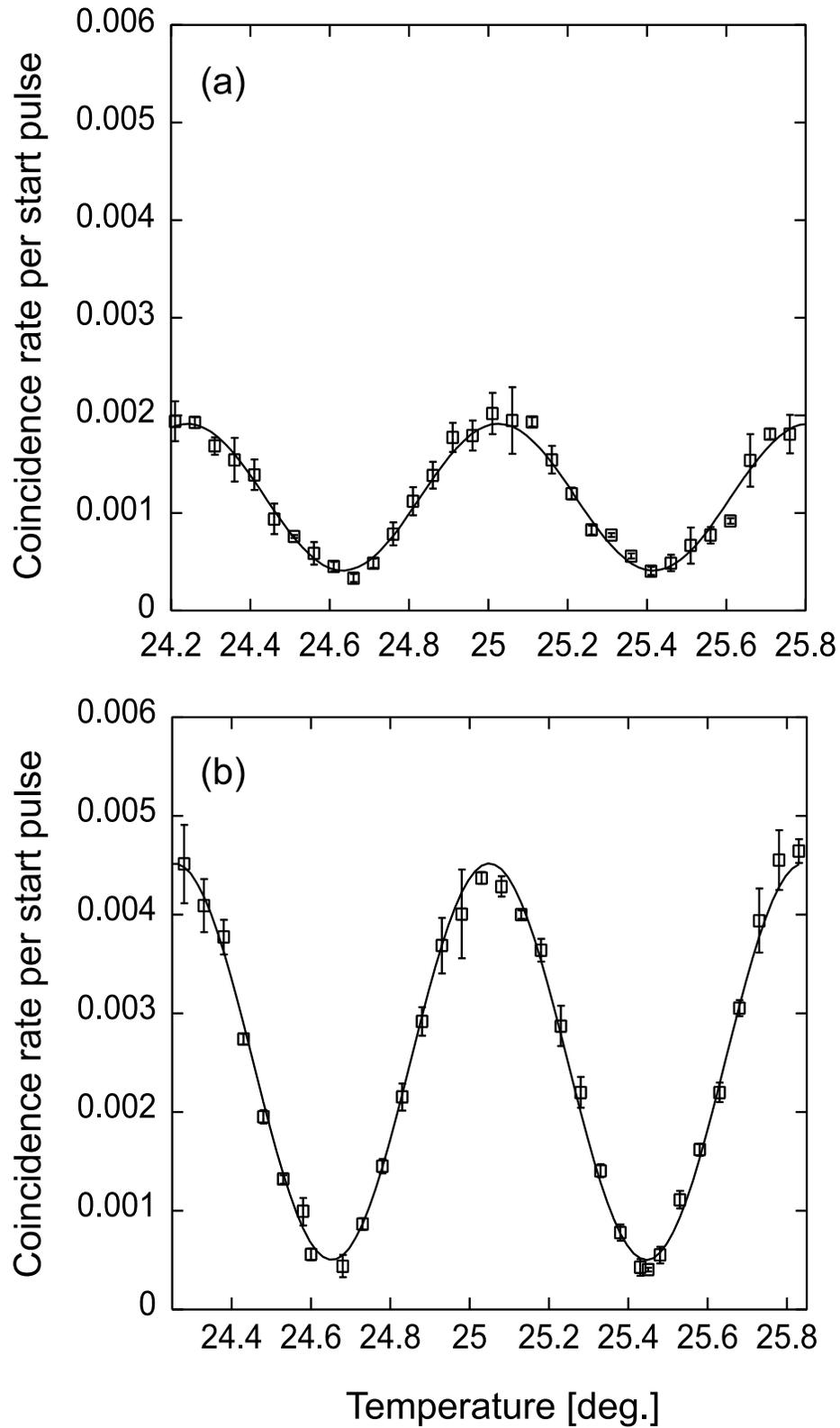}}

\caption{Two-photon interference fringes when the DSF was (a) at room temperature and (b) in liquid nitrogen.}
\label{fringe}

\end{figure}

First, I confirmed the effectiveness of fiber cooling for improving the visibility of two-photon interference without connecting 30-km DSF spools. I changed $\theta_i$ by changing the temperature of the interferometer for the idler, while fixing $\theta_s$, and recorded the coincidence counts. $\mu_s$ and $\mu_i$ were set at approximately 0.05 and 0.06, respectively, for both the cooled and uncooled experiments \cite{memo}. The average count rates of the signal and idler channels, respectively, were approximately 1500 and 1600 Hz throughout measurements. Without cooling the DSF, the visibility of the two-photon interference fringe was 64.7\%, which was obtained without removing the accidental coincidences (Fig. \ref{fringe} (a)). Fig. \ref{fringe} (b) shows the fringe when the DSF was cooled. The level of the minimum points of a fringe corresponds to the number of accidental coincidences, which is proportional to $\mu_s \mu_i$ as shown in Eq. (\ref{acc}). Because $\mu_s$ and $\mu_i$ were set at the same value for both measurements, the minimum points of both fringes were at almost the same level, as seen in Fig. \ref{fringe}. However, the peak level of the fringe increased significantly when the DSF was cooled. This implies that the number of noise photons is suppressed and so the portion of correlated photon pairs is effectively increased by cooling the DSF. As a result, the visibility increased to 80.0\% with the accidental coincidences included. 
The average number of correlated photon pairs per pulse $\mu_c$ can be estimated from the obtained visibilities and Eqs. (\ref{cor})-(\ref{v}). As a result, $\mu_c$ was $\sim$0.02 when the DSF was uncooled and $\sim$0.04 when cooled. Thus, it is experimentally confirmed that fiber cooling is effective for improving the visibility of a two-photon interference fringe.

\begin{figure}[hb]

\centerline{\includegraphics[width=.9\linewidth]{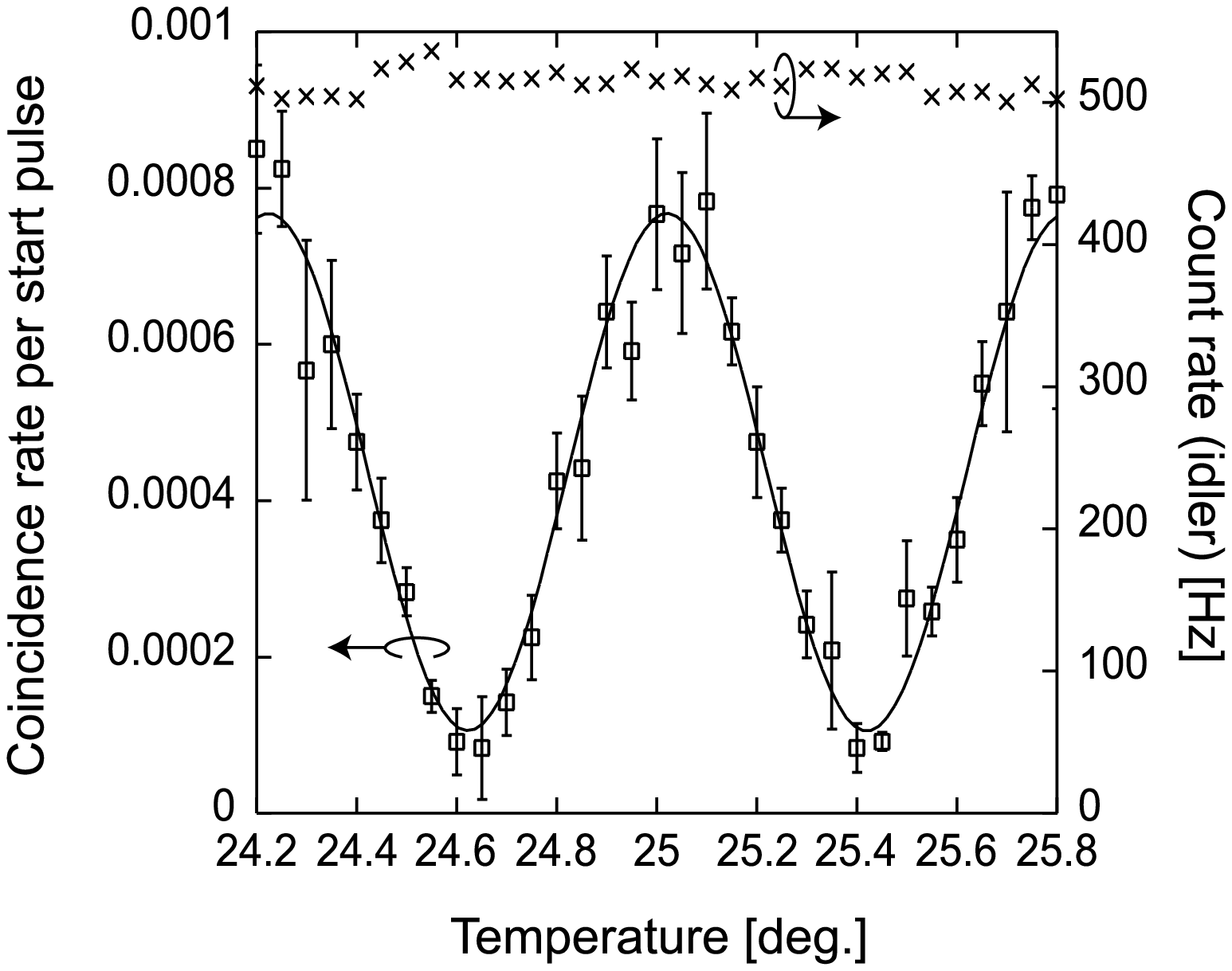}}

\caption{Two-photon interference fringe and idler count rate after transmission over 60-km fiber. Squares: coincidence rate per start pulse, x symbols: idler count rate.}
\label{60km}
\end{figure}

I then inserted a 30-km DSF spool between the bandpass filter and the interferometer in both the signal and idler arms, and undertook a two-photon interference experiment. The result is shown in Fig. \ref{60km}. $\mu_s$ and $\mu_i$ were again set at around 0.05 and 0.06, respectively. Squares show the coincidence rate per start pulse and x symbols show the idler count rate as a function of interferometer temperature. The average count rate for the signal was $\sim$430 Hz. The average coincidence rate at the peak of the fringe was as low as $\sim$0.3 Hz, which resulted in a long measurement time (the measurement shown in Fig. \ref{60km} took more than three hours to complete). The visibility of the fringe that includes accidental coincidences was 75.8\%, which exceeds the value associated with the violation of Bell's inequality ($\sim 71$\%). I would like to emphasize that the result presented here sets a new record for long-distance distribution of quantum entanglement.

In summary, I have reported the first experimental generation of time-bin entanglement using a cooled fiber. A significant improvement in the visibility was observed by cooling the DSF with liquid nitrogen. As a result, entangled photons were successfully distributed over 60-km (30 km x 2) fibers with a fair visibility of 75.8\%. The results show that a 1.5-$\mu$m band entanglement source based on a cooled fiber is a promising technology for realizing advanced quantum information systems over optical fiber networks. 

The author thanks K. Inoue for helpful comments and T. Honjo for help in making the measurement software. 
This work was supported in part by National Institute of Information and Communications Technology (NICT) of Japan.

\newpage

\begin{thebibliography}{99}
\bibitem{memo} Exactly speaking, $\mu_i$ was set at the same value in both cases, and $\mu_s$ for the cooled fiber was slightly smaller than that for the uncooled fiber. This is because, according to Eqs. (\ref{sss}) and (\ref{aaa}), the difference between Raman gain coefficients for the Stokes and anti-Stokes processes increases slightly as the fiber is cooled.
\end{thebibliography}
\end{document}